\documentclass{DISproc}

\begin{document}
\title{Global Analysis of Nuclear PDFs}

\author{{\slshape Daniel de Florian$^1$, Rodolfo Sassot$^1$, Marco Stratmann$^2$, Pia Zurita$^1$ }\\[1ex]
$^1$Departamento de F\'{\i}sica and IFIBA, FCEyN, Universidad de Buenos Aires, \\ Ciudad Universitaria, Pabell\'on\ 1 (1428) Buenos Aires, Argentina\\
$^2$Physics Department, Brookhaven National Laboratory, Upton, NY~11973, USA }

\contribID{xy}

\doi  

\maketitle

\begin{abstract}
We present a new global QCD analysis of nuclear parton distribution functions. 
In addition to the most commonly analyzed data sets for deep inelastic scattering 
of charged leptons off nuclei and Drell Yan di-lepton production, we include also 
measurements for neutrino-nuclei scattering as well as inclusive pion production 
in deuteron-gold collisions. The emerging picture is one of consistency, where 
universal nuclear modification factors for each parton flavor reproduce the main 
features of all data without any significant tension among the different sets.
\end{abstract}

\section{Motivation}

In the last few years, significant progress has been made in obtaining nuclear PDFs (nPDFs) 
from data. In addition to the theoretical improvements routinely used 
in modern extractions of free proton PDFs, such as the consistent implementation 
of QCD corrections beyond the LO \cite{ref:nds} and uncertainty estimates 
\cite{ref:hirai,ref:eps09}, the most recent determinations of nPDFs have also 
extended the types of data sets taken into account, moving towards truly global 
QCD analyses of nuclear effects \cite{ref:eps09,ref:schienbein,ref:paukkunen,Kovarik:2010uv}. 
The addition of novel hard probes to the fit does not only lead to better 
constrained sets of nPDFs and allows one to study the nuclear modification to the 
different parton species individually, but also tests the assumed process 
independence of nuclear effects.

The deep inelastic scattering (DIS) of charged leptons off nuclear targets
not only initiated all studies of nPDFs but still provides the best constraints 
on nuclear modifications for quark distributions. Upon combination with available 
data on Drell Yan (DY) di-lepton production off nuclear targets, a better 
discrimination between valence and sea quarks can be achieved. However, DIS and 
DY data only loosely constrain the nuclear modifications to the gluon density
because they cover a too small range in the hard energy scale $Q$.
To remedy this situation, data from BNL-RHIC for inclusive pion production in 
deuteron-gold (dAu) collisions have been included in the analysis of nPDFs performed 
in Ref.~\cite{ref:eps09}. Not surprisingly, these data have a significant impact 
in the determination of the gluon distribution. The corresponding nuclear 
modification for gluons turned out to be much more pronounced than in previous 
estimates and also much larger than those found for all the other partonic species.

Another promising avenue for significant improvements is neutrino induced DIS off 
iron and lead targets, with results available from NuTeV, CDHSW, and CHORUS \cite{deFlorian:2011fp}.
These data receive their importance from their discriminating power between 
nuclear modifications for quarks and antiquarks and have been included in a series
of analyses in Refs.~\cite{ref:schienbein,Kovarik:2010uv}.
Unexpectedly, the correction factors obtained from neutrino scattering data 
are found to differ significantly from those extracted with charged lepton probes 
\cite{ref:schienbein,Kovarik:2010uv}. At variance with these results, 
Ref.~\cite{ref:paukkunen} confronts the neutrino DIS cross sections with nPDFs 
obtained in \cite{ref:eps09} without any refitting and finds no apparent 
disagreement.

The novel global QCD analysis of nPDFs presented here \cite{deFlorian:2011fp} incorporates in a comprehensive 
way all of the above mentioned improvements and data sets. The resulting nPDFs at 
next-to-leading order accuracy supersede previous work presented in 
\cite{ref:nds}. We adopt a contemporary set of free nucleon PDFs \cite{Martin:2009iq} as our reference distribution to 
quantify modifications of PDFs in nuclei. As in \cite{Martin:2009iq}, we use a 
general mass variable flavor number scheme to treat charm and bottom 
quark contributions in our analysis. We use the Hessian method \cite{ref:hessian}
to estimate the uncertainties of the nuclear modification factors and examine 
critically their range of validity.

\section{Framework}

Throughout the analysis, we make the usual assumption that theoretical 
expressions for measured cross sections involving a nucleus factorize into 
calculable partonic hard scattering cross sections, identical to those used 
for processes involving free nucleons, and appropriate combinations of non-perturbative 
collinear parton densities and fragmentation functions.
The nPDFs $f^A_i(x,Q_0)$ at an initial scale $Q_0=1\,\mathrm{GeV}$ are related to proton 
distributions $f_i^p(x,Q_0)$ through a multiplicative nuclear modification factor 
$R^A_i(x,Q_0)$ as
\begin{equation}
 f^A_i(x,Q_0) = R^A_i(x,Q_0)\, f_i^p (x,Q_0)\;,
\label{eq:npdfdef}
\end{equation}
where $x$ is the usual DIS scaling variable for free nucleons.
Both valence quark distributions are assigned the same nuclear modification 
factor $R^A_v(x,Q^2_0)$ which we parametrize as
\begin{eqnarray}
R^A_v(x,Q^2_0)  =  \epsilon_1 \, x^{\alpha_v} (1-x)^{\beta_1} \;  
          (1 + \epsilon_2 (1-x)^{\beta_2})\, (1 + a_v  (1-x)^{\beta_3})\,.
\label{eq:rval}
\end{eqnarray}
We also assume that the light sea quarks and antiquarks share the same correction factor 
$R^A_s(x,Q^2_0)$. No significant improvement in the quality of 
the fit to data is found by relaxing this assumption. We choose another factor 
$R^A_g(x,Q^2_0)$ to parametrize medium effects for gluons.
An excellent description of the data is achieved by relating both $R_s^A$ and $R_g^A$ to
$R_v^A$ specified in Eq.~(\ref{eq:rval}), 
allowing only for a different normalization and modifications in the 
low-$x$ behavior. Hence we choose, without any loss in the quality of the fit, 
\begin{eqnarray}
 R^A_s(x,Q^2_0)  = R^A_v(x,Q^2_0) \,\frac{\epsilon_s}{\epsilon_1} \frac{1 + a_s x^{\alpha_s}}{a_s + 1}\;,\;\;\;\;\;
 R^A_g(x,Q^2_0)  = R^A_v(x,Q^2_0) \,\frac{\epsilon_g}{\epsilon_1} \frac{1 + a_g x^{\alpha_g}}{a_g + 1} \;.
\label{eq:rglue}
\end{eqnarray}
We note that the coefficients $\epsilon_1$ and  $\epsilon_2$ in Eq.~(\ref{eq:rval}) 
are fixed by charge conservation, and if we further constrain $\epsilon_s$ and 
$\epsilon_g$ to be equal, which, again, has no impact on the quality of the fit, 
$\epsilon_s$ is fixed by momentum conservation. The $A$ dependence of the remaining 
free parameters $\xi \in \{\alpha_v, \alpha_s,\alpha_g, \beta_1, \beta_2, \beta_3, a_v, a_s, a_g \}$ 
is parametrized in the usual way \cite{ref:nds} as $
\xi = \gamma_{\xi} + \lambda_{\xi} A^{\delta_{\xi}}$. 
The very mild $A$ dependence found for some of the $\xi$'s
allows us to further reduce the number of additional parameters 
by setting $\delta_{a_g}=\delta_{a_s}$ and 
$\delta_{\alpha_g}=\delta_{\alpha_s}$, leaving a total of
25 free parameters, which are obtained by a standard $\chi^2$
minimization, without artificial weights for certain data sets, i.e.\ $\omega_{i}=1$, 
and with statistical and systematic errors added in quadrature in $\Delta^2_i$:
\begin{equation}
\label{eq:chi2}
\chi^2 \equiv \sum_i \omega_{i} \,
\frac{ (d\sigma^{\mathrm{exp}}_i-d\sigma^{\mathrm{th}}_i)^2}
{\Delta^2_i} 
\end{equation} 

\section{Results}

\begin{wrapfigure}{r}{0.49\textwidth}
  \centering
	\vspace*{-1.0cm}
  \includegraphics[width=0.49\textwidth]{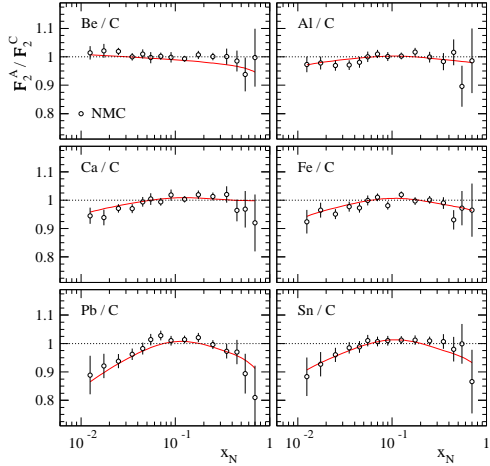}
  \vspace*{-0.8cm}
  \caption{Comparison to charged lepton DIS}
  \label{Fig:1}
\end{wrapfigure}
The total $\chi ^2$ for the optimum fit was found to be 1544.7 for
1579 data points ($\chi^2/d.o.f.=0.994$). All data sets are adequately reproduced, 
well within the nominal statistical range $\chi^2= n \pm \sqrt{2n}$ with $n$ the 
number of data.
More specifically, the partial contribution to $\chi^2$ of all the charged lepton 
DIS data amounts to 897.52 units for 894 data points, for neutrino DIS we find
488.20 units compared to 532 data points, DY observables amount to 90.72 units 
for 92 points, and pion production in $dAu$ collisions adds another 68.26 units 
to $\chi^2$ for 61 data points. 

In Figs.~1-3 we show some examples of the good agreement between the fit and charged
lepton DIS, neutrino DIS, and hadroproduction data, respectively;
see Ref.~\cite{deFlorian:2011fp} for details. The remarkable 
agreement with charged lepton DIS data, shown in Fig.~\ref{Fig:1}, is a common feature of all nPDFs analyses.
\begin{wrapfigure}{l}{0.49\textwidth}
  \centering
	\vspace*{-0.8cm}
  \includegraphics[width=0.49\textwidth]{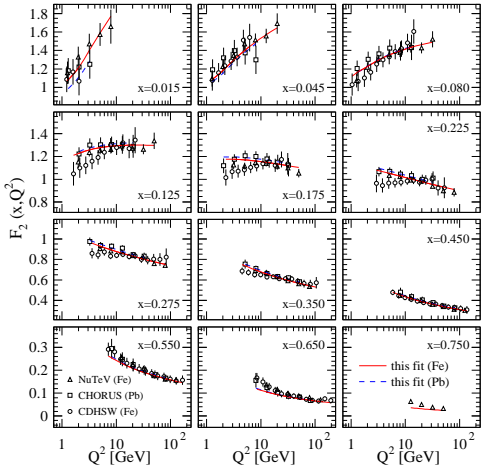}
  \vspace*{-0.8cm}
  \caption{Comparison to neutrino DIS data}
  \label{Fig:2}
\end{wrapfigure}

Neutrino DIS data for the averaged structure function $(F_2^{\nu A}+F_2^{\bar{\nu}A})/2$ are well reproduced 
within the experimental uncertainties both in shape and in magnitude, see Fig.~\ref{Fig:2}.
The only noticeable exception are the CDHSW data 
at $Q^2$ values below $10\,\mathrm{GeV}^2$ where they exhibit a rather different slope 
than the other data. In fact, in this $Q^2$ region it appears
to be impossible to simultaneously fit all data sets equally well, suggesting some
systematic discrepancy among the different neutrino data which needs to be investigated
further. Data for the averaged structure function $F_3$ are also well described 
by our fit \cite{deFlorian:2011fp}.

In general, results from $dAu$ collisions are significantly less straightforward to 
interpret in terms of nuclear modification factors. 
Each value of $p_T$ samples different fractions of the contributing
partonic hard scattering processes, integrated over a range of $x$.
Furthermore, since $p_T$ sets the magnitude for the factorization scale, 
the ratios reflect also the energy scale dependence of the effects. Apart 
from the nuclear modifications of parton densities, accounted for by the nPDFs,
the cross sections are in principle also sensitive to medium induced effects
in the hadronization process. 

Assuming factorizability for a given nucleus, such final-state effects can be absorbed into effective 
nuclear parton-to-hadron fragmentation functions (nFFs). The solid lines in Fig.~3 
represent the result of our best fit of nPDFs using the nFFs of Ref.~\cite{ref:nff}. 
The fit follows well the rise and fall of the ratio at small and high $p_T$,
respectively, but falls somewhat short in reproducing the enhancement found at 
medium $p_T$. Owing to the large experimental uncertainties,
\begin{wrapfigure}{r}{0.49\textwidth}
  \vspace*{-0.8cm}
  \includegraphics[width=0.49\textwidth]{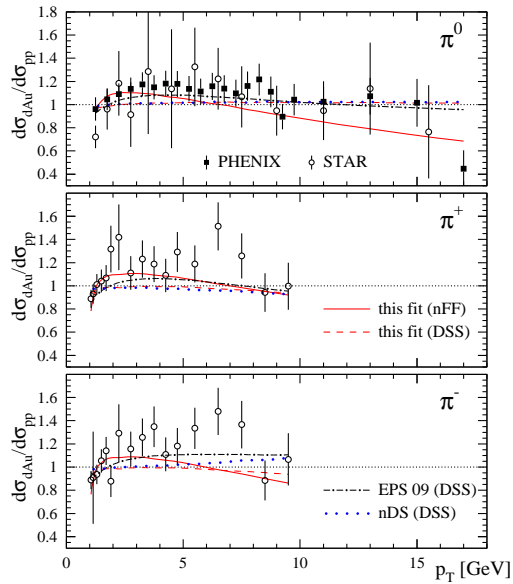}
  \vspace*{-0.8cm}
  \caption{Pion production in dAu collisions}
  \label{Fig:3}
\end{wrapfigure}
the $\chi^2$ for this 
subset of data is nevertheless good, $\chi^2_{dAu}/n=1.12$, in 
particular, if compared to the outcome of an otherwise similar fit using vacuum FFs \cite{deFlorian:2007aj} 
where $\chi^2_{dAu}/n=1.37$.
Data for $\pi^0$ yields in $dAu$ collisions were first incorporated by 
EPS \cite{ref:eps09} and found to provide a vital constraint 
on $R_g^{Au}$. At variance with our approach, the authors in \cite{ref:eps09} 
disregard any medium modifications in the hadronization and assign a large weight
$\omega_{dAu}$ in Eq.~(\ref{eq:chi2}), 
which drives their observed large nuclear modifications of the gluon density. 
Our $R_g^{Au}$ exhibits only moderate nuclear corrections.

Uncertainties in the extraction of our nPDFs have been estimated 
with the Hessian method \cite{ref:hessian} for a tolerance criterion of
$\Delta \chi^2=30$ and found to be rather large \cite{deFlorian:2011fp}, 
in particular, when compared to the present knowledge of free proton PDFs.
As always, these estimates are only trustworthy in the region constrained by data,
i.e., $x> 0.01$. In particular, prompt photon and DY di-lepton production
in $dAu$ and $pPb$ collisions at RHIC and the LHC, respectively, will help to 
further constrain nPDFs in the future; see, e.g., Ref.~\cite{deFlorian:2011fp}
for some quantitative expectations.

\section*{Acknowledgements}
M.S.\ acknowledges support by the U.S.\ DOE under contract 
no.\ DE-AC02-98CH10886. This work was supported in part by CONICET, ANPCyT, 
UBACyT, and the E.U.\ under grant no.\ PITN-GA-2010-264564 (LHCPhenoNet).


{\raggedright
\begin{footnotesize}




\begin{thebibliography}{99}

%
\bibitem{ref:nds}
  D.~de Florian and R.~Sassot, Phys. Rev.  {\bf D69}  (2004) 074028.
%
\bibitem{ref:hirai}
  M.~Hirai, S.~Kumano, and T.~H.~Nagai,
  Phys. Rev.  {\bf  C76}  (2007) 065207.
%
\bibitem{ref:eps09}
  K.~J.~Eskola, H.~Paukkunen, and C.~A.~Salgado,
  JHEP {\bf 0904} (2009)  065.
%
\bibitem{ref:schienbein}
  I.~Schienbein {\it et al.},
  Phys. Rev.  {\bf D77}  (2008) 054013;
  {\bf D80} (2009)  094004.
%
\bibitem{ref:paukkunen} 
  H.~Paukkunen and C.~A.~Salgado,
  JHEP\ {\bf 1007}  (2010)  032.
%
\bibitem{Kovarik:2010uv}
  K.~Kovarik {\it et al.},
  Phys. Rev. Lett.  {\bf 106} (2011) 122301.
%
\bibitem{deFlorian:2011fp}
  D.~de Florian, R.~Sassot, M.~Stratmann, and P.\ Zurita,
ÊÊarXiv:1112.6324 and references therein.
ÊÊ
%
\bibitem{Martin:2009iq}
  A.~D.~Martin, W.~J.~Stirling, R.~S.~Thorne, and G.~Watt,
  Eur. Phys. J.  {\bf C63}  (2009) 189.
%
\bibitem{ref:hessian}
  J.~Pumplin {\it et al.},
  Phys. Rev. {\bf D65} (2001) 014011;
  Phys. Rev.  {\bf D65} (2001) 014013.
%
\bibitem{ref:nff}
  R.~Sassot, M.~Stratmann, and P.~Zurita,
  Phys. Rev. {\bf D81} (2010) 054001.  
%
\bibitem{deFlorian:2007aj} 
  D.~de Florian, R.~Sassot, and M.~Stratmann,
  Phys.\ Rev.\ {\bf D75}  (2007) 114010.

\end{thebibliography}
\end{footnotesize}
}


\end{document}